\documentclass{emulateapj}

\newcommand{\be}{\begin{eqnarray}}
\newcommand{\ee}{\end{eqnarray}}

\shorttitle{EM Emission post Black Hole-Neutron Star Mergers}
\shortauthors{Zhong, Dai, \& Deng}

\begin{document}
\title{Electromagnetic Emission post Spinning Black Hole-Magnetized Neutron Star Mergers}
\author{Shu-Qing Zhong}
\affil{School of Astronomy and Space Science, Nanjing University, Nanjing 210093, China}
\affil{Key laboratory of Modern Astronomy and Astrophysics (Nanjing University), Ministry of Education, Nanjing 210093, China}
\author{Zi-Gao Dai}
\affil{School of Astronomy and Space Science, Nanjing University, Nanjing 210093, China; dzg@nju.edu.cn}
\affil{Key laboratory of Modern Astronomy and Astrophysics (Nanjing University), Ministry of Education, Nanjing 210093, China}
\author{Can-Min Deng}
\affil{Department of Astronomy, School of Physical Sciences, University of Science and Technology of China, Hefei, Anhui 230026, China; dengcm@ustc.edu.cn}
\affil{CAS Key Laboratory for Research in Galaxies and Cosmology, Department of Astronomy, University of Science and Technology of China, Hefei 230026, Anhui, China}
\affil{School of Astronomy and Space Science, University of Science and Technology of China, Hefei, Anhui 230026, China}

\begin{abstract}
For a binary composed of a spinning black hole (BH) (with mass $\gtrsim 7M_\odot$) and a strongly magnetized neutron star (NS) (with surface magnetic field strength $B_{\rm S,NS}\gtrsim10^{12}$\,G and mass $\sim 1.4M_\odot$), the NS as a whole will possibly eventually plunge into the BH. During the inspiral phase, the spinning BH could be charged to the Wald charge quantity $Q_{\rm W}$ until merger in an electro-vacuum approximation. During the merger, if the spinning charged BH creates its own magnetosphere due to an electric field strong enough for pair cascades to spark, the charged BH would transit from electro-vacuum to force-free cases and could discharge in a time $\gtrsim1~{\rm ms}$. As the force-free magnetosphere is full of a highly conducting plasma, the magnetic flux over the NS's caps would be retained outside the BH's event horizon under the frozen-in condition. Based on this scenario, we here investigate three possible energy-dissipation mechanisms that could produce electromagnetic (EM) counterparts in a time interval of the BH's discharge post a BH-NS merger-induced gravitational wave event: (1) magnetic reconnection at the BH's poles would occur, leading to a millisecond bright EM signal, (2) a magnetic shock in the zone of closed magnetic field lines due to the detachment and reconnection of the entire BH magnetic field would probably produce a bright radio emission, e.g., a fast radio burst, and (3) the Blandford-Znajek mechanism would extract the BH's rotational energy, giving rise to a millisecond-duration luminous high-energy burst. We also calculate the luminosities due to these mechanisms as a function of BH's spin for different values of $B_{\rm S,NS}$.
\end{abstract}


\section{Introduction}
\label{sec:introduction}

Eleven gravitational wave (GW) events have been observed from ten black hole-black hole (BH-BH) mergers and one neutron star-neutron star (NS-NS) merger in the observing run 1 \& 2 (O1 \& O2) of the Advanced LIGO and Virgo detectors \citep{ligo2018}, including GW150914 \citep{abb2016}. The first NS-NS merger, GW170817, jointly detected with GW and electromagnetic (EM) radiation, bas been a watershed event \citep{abb2017a,abb2017b,abb2017c}, which was accompanied by a weak short gamma-ray burst (GRB) 170817A, an early multi-wavelength kilonova AT2017gfo \citep{cou2017,villar2017}, and a late broadband afterglow \citep{von2017,gold2017,sav2017,hall2017,troja2017,lyman2018,mar2018,piro2019}. However, what has not yet been formally observed is a BH-NS merger, even though a candidate GW event S190426c in O3 running was presented \citep{ligo2019}, in which no evidence for follow-up EM counterparts was uncovered recently \citep{hoss2019}.

During a BH-NS merger, theoretically, the NS would be effectively tidally-disrupted if the tidal radius ($R_{\rm tidal}$) is larger than the innermost stable circular orbit (ISCO, $R_{\rm ISCO}$) of the BH, $R_{\rm tidal}>R_{\rm ISCO}$ (or the BH-NS mass ratio, $q\equiv M_{\rm BH}/M_{\rm NS}\lesssim5$) \citep{shi2009,bart2013}. The disrupted materials would be left behind outside the final BH remnant but some of them would fall back, generating a short GRB \citep{pacz1991} and an afterglow \citep{sari1998}, accompanied by a radioactively-powered kilonova, just as in the case of an NS-NS merger \citep{lp1998,fm2016,metz2017}. If the mass ratio $q\gtrsim5$, however, it is generally thought that the NS would plunge into the BH as a whole, leaving behind neither any material outside the BH remnant nor any significant EM emission.

Nevertheless, several possible scenarios have been proposed to produce EM counterparts prior to plunging BH-NS mergers. First, a BH battery power due to a spinning BH-magnetized NS (or non-spinning BH-spinning magnetized NS) binary inspiral behaving analogously to a unipolar inductor can serve as an energy source to generate EM emission such as a fast radio burst (FRB) or a relativistic fireball as long as the NS is strongly magnetized \citep{mcw2011,do2013,ming2015,do2016}. Second, if at least one of the members in a BH-NS binary is charged, such a binary could make an FRB or a short GRB through electric dipole radiation and magnetic dipole radiation \citep{zhang2016,zhang2019a,deng2018}. Third, if the BH is highly spinning and immersed in a strongly  magnetic field of the NS, the BH could be charged through the Wald's mechanism \citep{wald1974} as a BH pulsar \citep{levin2018}, and could create EM counterparts such as short X/$\gamma$-ray signals due to  magnetic reconnection in addition to electric dipole radiation and magnetic dipole radiation \citep{dai2019}, as in the cases of binary BH mergers \citep{fra2018} and binary NS mergers \citep{wang2018}.

Moreover, it has been argued that EM counterparts could be produced after a plunging BH-NS merger. If the binary includes a non-spinning BH and a spinning magnetized NS, a ``blitzar'' FRB would occur via a magnetic shock because of the ``no-hair'' theorem \citep{fr2014,ming2015}. On the other hand, for a spinning BH-magnetized NS binary, the ``no-hair'' theorem could not be applicable formally and the magnetic flux at the plunging NS polar caps would be retained in the highly conducting plasma outside the BH's event horizon during the resistive timescale of the magnetosphere \citep{lyu2011}, so that a type of afterglow post BH-NS merger would be powered via the Blandford-Znajek (BZ) mechanism \citep{bz1977,do2016}.

During the inspiral of a spinning massive BH-magnetized NS binary, the BH could be charged because it is immersed in the magnetosphere of the NS, so that this system would become a spinning charged BH-magnetized NS binary. Its pre-merger energy-dissipation mechanisms such as magnetic reconnection, electric dipole radiation, and magnetic dipole radiation could give rise to EM emission \citep{dai2019}. In this paper, we investigate post-merger possible mechanisms to generate EM counterparts that are suitable for a spinning massive BH-magnetized NS binary in which the BH is charged via the Wald process, based on the previously-studied scenarios in some binary systems, e.g., a spinning charged BH-magnetized NS binary premerger, a non-spinning BH-spinning magnetized NS binary postmerger, and a spinning BH-magnetized NS binary postmerger. Ultimately, we find that three possible mechanisms such as magnetic reconnection, magnetic shock, and BZ mechanism are interesting. This paper is organized as follows. In \S\,2 we briefly illustrate the schematic picture of these three mechanisms. In \S\,3, we calculate the emission luminosities due to the mechanisms. Finally, we present a discussion and summary in \S\,4.

\section{Processes Post a Magnetized NS Plunging into a Spinning BH}
\label{sec:NS-BH}

For convenience of numerical calculations, we show the quantities in cgs units. We use a subscript ``NS'' to indicate the quantities of an NS (those without subscript ``NS'' are the BH's). For a spinning BH-magnetized NS binary in an electro-vacuum approximation, as shown in \cite{levin2018}, the BH could not only be charged stably but also maintain the Wald's charge quantity $Q=Q_{\rm W}$ until the NS plunges into the BH, even though EM emission during the charging process and continual flux of charges within the BH-NS system occurs. During this plunging period, as the separation between BH and NS reaches the NS's radius ($r\rightarrow R_{\rm NS}$), Wald's maximal charge quantity of the BH is
\be
	Q_{\rm W,max} & \simeq & \frac{2G}{c^3}J\times B_{\rm S,NS}
    =\frac{2G^2}{c^4}aM^2B_{\rm S,NS} 	\nonumber\\
    &=&4.4\times10^{24}a\left(\frac{M}{10M_{\odot}}\right)^{2}\frac{B_{\rm S,NS}}{10^{12}{\rm G}}~{\rm e.s.u.},
	\label{eq:Q_W,max}
\ee
where $J$, $a=Jc/GM^2$, $M$, and $B_{\rm S,NS}$ are the BH's angular momentum, dimensionless spin parameter, mass, and the NS's surface magnetic dipole field strength, respectively.
The Wald charge would give rise to a magnetic dipole field of the BH
whose strength at any radius $r$ is \citep{levin2018,dai2019}
\be
B_{\rm W}&=&\frac{JQ_{\rm W,max}}{Mcr^3}=\frac{2G^3}{c^6}a^2M^3B_{\rm S,NS}r^{-3}	\nonumber\\
&=&6.5\times10^{12}a^{2}\left(\frac{M}{10M_{\odot}}\right)^{3}   \nonumber\\
&\times&\left(\frac{B_{\rm S,NS}}{10^{12} \mathrm{G}}\right)\left(\frac{r}{10^{6}\mathrm{cm}}\right)^{-3}~\mathrm{G}.
\label{eq:B_W}
\ee
During the merger, if the spinning charged BH creates its own magnetosphere due to an electric field strong enough for pair cascades to spark, the charge BH would transit from electro-vacuum to force-free cases. This does not mean that the BH would completely discharge immediately. As shown in \cite{py2019}, the duration of BH discharge may be larger than the light crossing time $\sim1$ ms. However, if the resistive time of a force-free magnetosphere is considered \citep{lyu2011}, the duration of BH discharge could be longer. This duration is comparable to the duration of the mechanisms that we discuss below. So these mechanisms would occur even if the BH discharge takes place in the post-merger force-free case.

In the scenario discussed above, on one hand, no matter whether the BH spin aligns or anti-aligns with the NS magnetic axis before the NS is swallowed, the BH's magnetic dipole field created from the spinning electric charge inside the BH event horizon is always anti-parallel to that of the NS \citep{dai2019}, so a magnetic reconnection event could occur in an interacting field zone close to the BH equatorial plane initially \citep{dai2019}, as in the case of a binary NS premerger \citep{wang2018}. During the NS plunging into the BH, the NS's closed magnetic field lines would be swallowed together with the NS, while the open field lines which originally connect the NS's polar surface to infinity also connect the BH's event horizon to infinity because of the frozen-in condition of a highly conducting plasma, which is akin to the cases in a rotating NS collapsing to a slowly balding BH \citep{lyu2011} or in a strongly-magnetized NS-Kerr BH binary postmerger \citep{do2016}. Therefore, the magnetic reconnection zone close to the BH's equatorial plane premerger in \cite{dai2019} would be transferred to the BH polar regions postmerger, in which regions the open field lines of the BH would reconnect with those of the NS.

\begin{figure}
\center
\includegraphics[width=0.5\textwidth, angle=0]{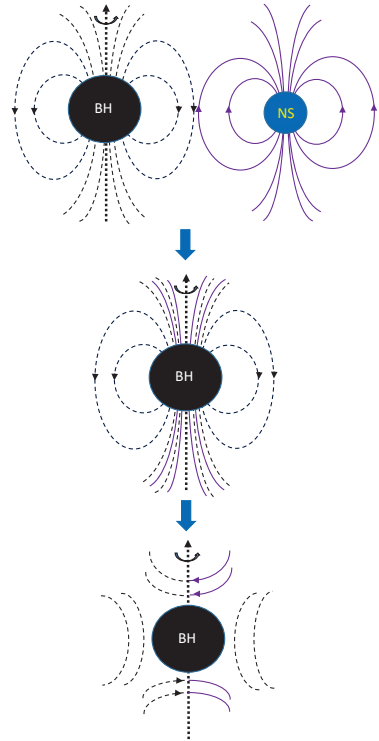}
\caption{The schematic picture of a plunging spinning BH-magnetized NS binary premerger and postmerger (for convenience, we just show some of  the magnetic field lines of the NS in the top panel. In reality, the farthest closed magnetic field lines of the NS should already thread the BH since the BH immersed in the NS's magnetosphere could be charged). Prior to the merger, the spinning BH would not only be charged stably up to the Wald charge $Q_{\rm W}$ but also maintain it up to its maximum quantity in an electro-vacuum approximation. During the merger, if the spinning charged BH creates its own magnetosphere due to an electric field strong enough for pair cascades to spark, the charge BH would transit from electro-vacuum to force-free cases. Furthermore, during the NS plunging into the BH, the NS's closed magnetic field lines would be swallowed together with the NS, while the open field lines which originally connect the NS's polar surface to infinity also connect the BH's event horizon to infinity because of the frozen-in condition of a highly conducting plasma. In this scenario, on one hand, the open magnetic field lines of the BH at polar regions would reconnect with those of the NS because the BH magnetic dipole field lines are always anti-parallel to those of the NS. On the other hand, the BH's discharge might create a magnetic shock due to the detachment and reconnection of the entire magnetic field of the BH, leading to the magnetosphere plasma dissipation. In addition, the third process---BZ mechanism may occur to extract the BH's rotational energy.}
\label{fig:schematic}
\end{figure}

On the other hand, the BH discharge process, just like the magnetosphere collapse due to the magnetosphere instability proposed in \cite{liu2016}, would in principle result in an entire magnetic field detachment and reconnection, which generates a strong magnetic shock at the speed of light to sweep up the magnetosphere plasma, similar to the mechanism for a supramassive the NS collapsing to a BH\footnote{A strong magnetic shock wave is seen in 3D resistive MHD simulations on the collapse of non-rotating NSs \citep{dion2013}.}, proposed by \cite{fr2014}. In addition, the third mechanism---BZ mechanism may occur to extract the BH's rotational energy, as in the case of a strongly-magnetized NS-Kerr BH binary postmerger \citep{do2016}. The schematic picture for these processes is shown in Figure \ref{fig:schematic}.

\section{Mechanisms of EM Emission Postmerger}
\label{sec:Mechanisms}

\subsection{Magnetic Reconnection}
\label{subsec:Magnetic Reconnection}

The magnetic reconnection would be transferred from the region close to the BH's equatorial plane premerger to the polar regions postmerger. The energy of each pole at radius $r-r+dr$,
\be
dE_{\mathrm{REC}}(r)=\frac{B_{\rm W}^2}{8 \pi}dV_{p}(r),
\label{eq:E_REC}
\ee
can be dissipated, where $dV_{p}(r)$ is the volume in an interval of $r$ to $r+dr$.
Based on Eqs. (\ref{eq:Q_W,max}) and (\ref{eq:B_W}), the magnetic field at the BH's horizon radius \citep{misner1973}
\be
R_{\rm H}&=&\frac{GM}{c^2}+\sqrt{\frac{G^2M^2}{c^4}-\frac{G^2M^2}{c^4}a^2-\frac{GQ_{\rm W,max}^2}{c^4}} \nonumber\\
&\approx& \frac{GM}{c^2}+\sqrt{\frac{G^2M^2}{c^4}-\frac{G^2M^2}{c^4}a^2},
\label{eq:R_H}
\ee
can reach $10^{11}~{\rm G}$ for the BH mass $\sim 10M_{\odot}$, the NS's surface magnetic field strength $\sim 10^{12}~{\rm G}$, and the spin parameter $a\sim0.6$. Hence, for a relatively low BH spin $a<0.6$, the strength of BH's horizon magnetic field is much smaller than the surface magnetic field of the NS. In this case, the BH's open magnetic field lines can only reconnect a small part of the open magnetic field lines of the NS.
We next calculate the volume $dV_{\rm p}(r)$ with an analogy to that of a pulsar
\be
dV_{p}(r)=A(r)dr,
\label{eq:Volume}
\ee
where $A(r)=\pi r^2 \sin^2\theta\sim\pi r^2 R_{\rm H}/R_{\rm LC}$, and $R_{\rm LC}$ is the light cylinder radius of the BH.
Here any dipole magnetic field line satisfies $r'=r'_{\rm max}\sin^2\theta$, where $r'\sim R_{\rm H}$ and $r'_{\rm max}\sim R_{\rm LC}$ are adopted \citep{piro2011,wang2018}.

\begin{figure}
\plotone{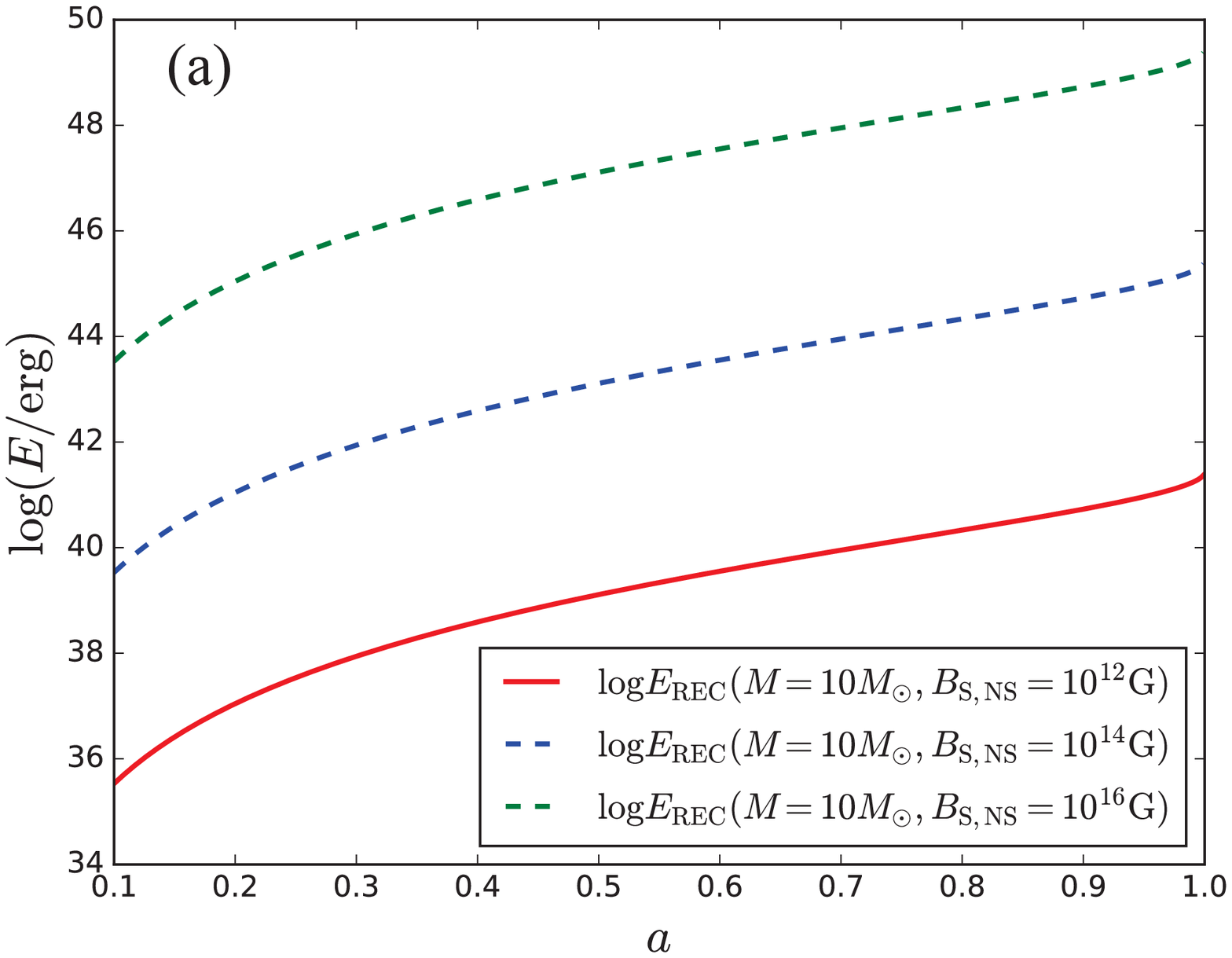}
\plotone{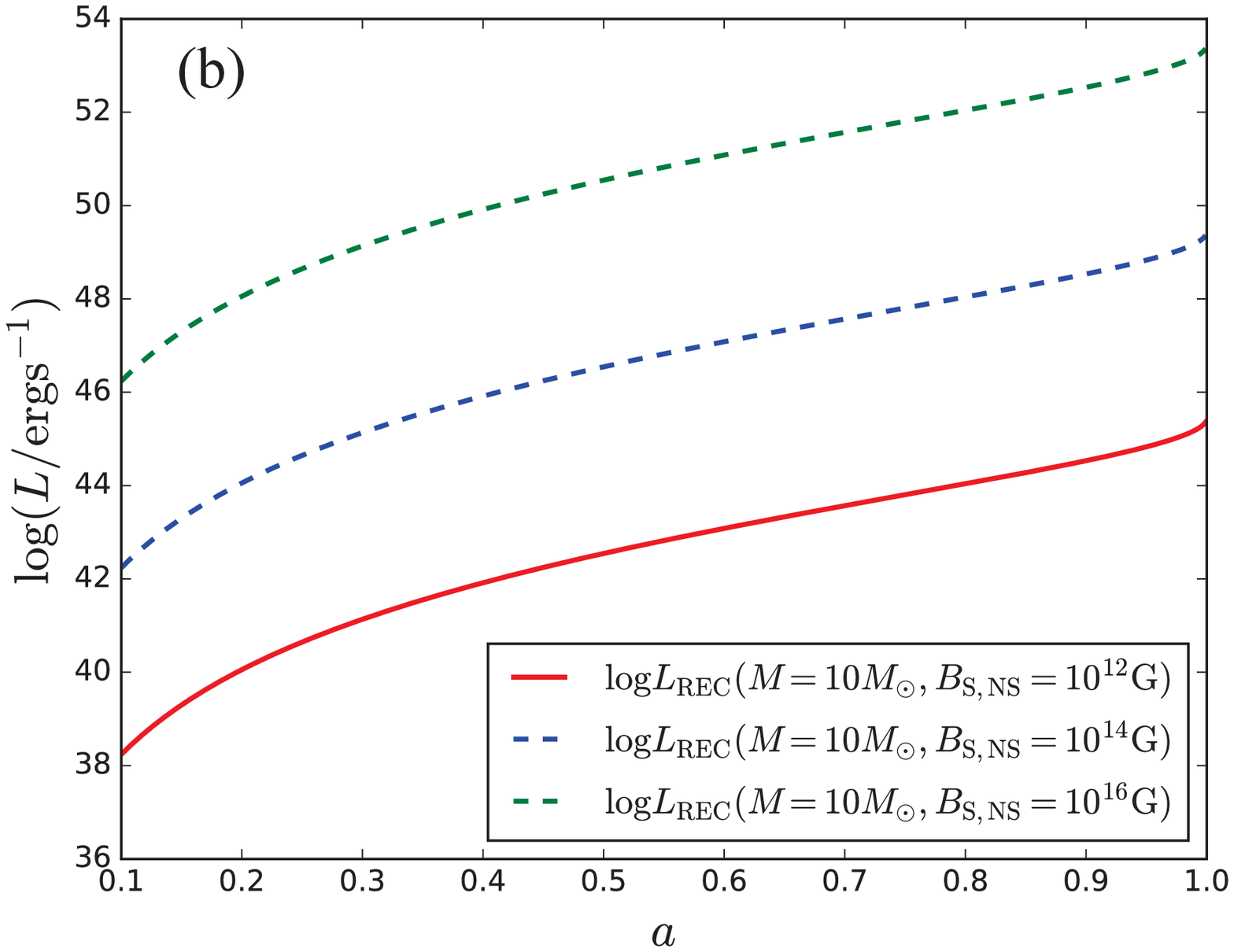}
\caption{Panels {\bf a} and {\bf b}: The magnetic reconnection energy $E_{\rm REC}$ at each BH's pole and the corresponding luminosity $L_{\rm REC}$ as a function of BH dimensionless spin parameter $a$ for different $B_{\rm S,NS}$.}
\label{fig:REC-a}
\end{figure}

The light cylinder radius can be written as
\be
R_{\rm LC}=c/\Omega,
\label{eq:R_LC}
\ee
where the angular velocity is approximated as that close to the event horizon \citep[see Eq. 5 of][]{liu2016}
\be
\Omega=\frac{GMa}{c[R_{\rm H}^2+(GMa/c^2)^2]}.
\label{eq:Omega}
\ee
From Eqs. (\ref{eq:Q_W,max}), (\ref{eq:B_W}), (\ref{eq:E_REC}), (\ref{eq:Volume}),
and (\ref{eq:R_LC}),
one can obtain the total magnetic reconnection energy
for each polar region
\be
E_{\rm REC}&=&\int_{R_{\rm H}}^{R_{\rm LC}}\frac{B_{\rm W}^2}{8 \pi}\frac{R_{\rm H}}{R_{\rm LC}}\pi r^2 dr \nonumber \\
&\approx& \frac{G^6a^4M^6B_{\rm S,NS}^2}{6c^{12}}\frac{1}{R_{\rm H}^2R_{\rm LC}},
\label{eq:E_tot_rec}
\ee
which can be as a function of the spin parameter $a$
for different NS's surface magnetic fields $B_{\rm S,NS}=10^{12}, 10^{14}$,
and $10^{16}~{\rm G}$, and BH mass $M=10M_{\odot}$, displaying in Panel a of Figure \ref{fig:REC-a}. The magnetic reconnection energy rises with increasing the NS's surface magnetic field and the BH spin.

If the magnetic reconnection dissipation timescale is assumed to be the Alfv\'en propagation timescale $t_{\rm{A}}$,
this timescale can be expressed as the ratio of the light cylinder of the BH ($R_{\rm LC}$) to
the Alfv\'en speed ($v_{\rm{A}}$),
\be
t_{\mathrm{A}}=\frac{R_{\mathrm{LC}}}{v_{\mathrm{A}}},
\label{eq:t_A}
\ee
where $v_{\rm{A}}$ can be generally expressed by \citep[see Eq. 2.11a in][]{punsly1998}
\be
v_{\rm A}=\frac{U_{\rm A}}{\gamma_{\rm A}}=\frac{B_{\rm W}}{\gamma_{\rm A}\sqrt{4\pi n_e\mu}}c,
\label{eq:v_A}
\ee
where $n_e$ is the particle number density in the BH's magnetosphere, $U_{\rm A}$ is the pure Alfv\'en speed and
$\gamma_{\rm A}=\frac{1}{\sqrt{1-v_{\rm A}^2/c^2}}$,
and $\mu\sim10^{-6}$ is the plasma specific enthalpy in units of the electron's rest mass energy\footnote{The quantity $\mu$ is calculated based on the assumption that the plasma is cold. $\mu$ is composed of the energy density $\rho$ and the pressure $P$, where $\rho$ consists of the corresponding energy of the rest mass ($m_{e}c^{2}$) and the internal energy $I$ per particle. The internal energy per particle ($I$) is basically the random kinetic energy for an ordered EM field as done here. The field-aligned acceleration produces an ordered field-aligned motion and does not produce a random kinetic energy so that it does not affect the internal energy per particle. The collision rate affects $P$ and $I$, in a relatively low density approximation that both $P$ and $I$ are small compared to $m_{e}c^{2}$ \citep[see Eq. 2.13 in][]{punsly2001}.}.
The particle number density $n_e$ in the KNBH's magnetosphere is unlikely to be calculated accurately due to complexity of the magnetosphere, even though a KNBH's magnetosphere model with a magnetically dominated plasma has been explored in \cite{punsly1998}. In his model, the particle number density is nonlinear and uncertain no matter for the MHD wind zone and the zone of closed magnetic field lines of the magnetosphere, even though it has been simplified under several assumptions. This may be why \cite{liu2016} did not adopt this magnetosphere model to calculate the particle number density for a KNBH. Instead, \cite{liu2016} took $n_e$ to be a factor $k$ times the Goldreich-Julian density \citep{gj1969} \citep[as done in][]{fr2014}
\be
n_e=k\frac{B_{\rm W}\Omega}{2\pi ce},
\label{eq:n_e}
\ee
where $k$ is uncertain but it is assumed to be typically of order unity here, $k\sim O(1)$. In the following, we always take $k=1$. This assumption seems to be reasonable based on the model of \cite{punsly1998}, in which the particle number density has a basically similar but more complicated form associated with the magnetic field strength and angular velocity of the BH no matter for the MHD wind zone or the zone of closed magnetic field lines of magnetosphere \citep[see Eqs. 4.12a and A.12 in][]{punsly1998}.
From Eqs. (\ref{eq:v_A}) and (\ref{eq:n_e}), the Alfv\'en speed can be also written as
$v_{\rm A}=\frac{B_{\rm W}^{1/2}\Omega^{-1/2}c}{\gamma_{\rm A}\sqrt{2k\mu/ce}}$, which implies the Alfv\'en speed close to $R_{\rm LC}$ is smaller than that close to the BH horizon. Thus, we use the Alfv\'en speed close to $R_{\rm LC}$ to represent the Alfv\'en speed of the magnetosphere.
Hence, if $B_{\rm S,NS}=10^{12}~{\rm G}$, $M=10M_{\odot}$, and $a\sim0.1-1.0$, then we obtain $B_{\rm W}\ge10^5~{\rm G}$ for the whole magnetosphere ranging from $R_{\rm H}$ to $R_{\rm LC}$ and $v_{\rm{A}}$ basically equals to the speed of light $c$.
Moreover, the stronger NS's magnetic field corresponds to the higher Alfv\'en speed, as seen from the expression of $v_{\rm A}$ depending on $B_{\rm W}$ and Eq. (\ref{eq:B_W}).
Using $a=0.1-1.0$ and $M=10M_{\odot}$, we further get
the BH magnetosphere size $R_{\rm LC}\sim 3\times10^6-6\times10^7~{\rm cm}$ by combining Eqs. (\ref{eq:R_LC}) and (\ref{eq:Omega}), so the Alfv\'en propagation timescale $t_{\rm{A}}\sim0.1-2~{\rm ms}$.
Moreover, in terms of Eqs. (\ref{eq:Q_W,max}), (\ref{eq:B_W}),
(\ref{eq:R_LC}), (\ref{eq:Omega}), (\ref{eq:E_tot_rec}), (\ref{eq:t_A}), and (\ref{eq:v_A}),
the total magnetic energy in each polar region of BH would be dissipated
as a luminosity $L_{\mathrm{REC}}=E_{\mathrm{REC}}/t_{\rm A}$, illustrated in Panel b of Figure \ref{fig:REC-a}.
The luminosity increases with increasing the BH spin and NS's surface magnetic field strength. Furthermore, it is comparable to the luminosity in a magnetic reconnection event close to the equatorial plane from the magnetospheres' contact of two NSs with a surface magnetic field $10^{12}~{\rm G}$ premerger \cite[in case 1 in Fig. 2 of][]{wang2018}, only if the BH spin $a\gtrsim0.6$.

\subsection{Magnetic Shock}
\label{subsec:Magnetic Shock}

\begin{figure}
\includegraphics[width=0.4\textwidth, height=0.18\textheight, angle=0]{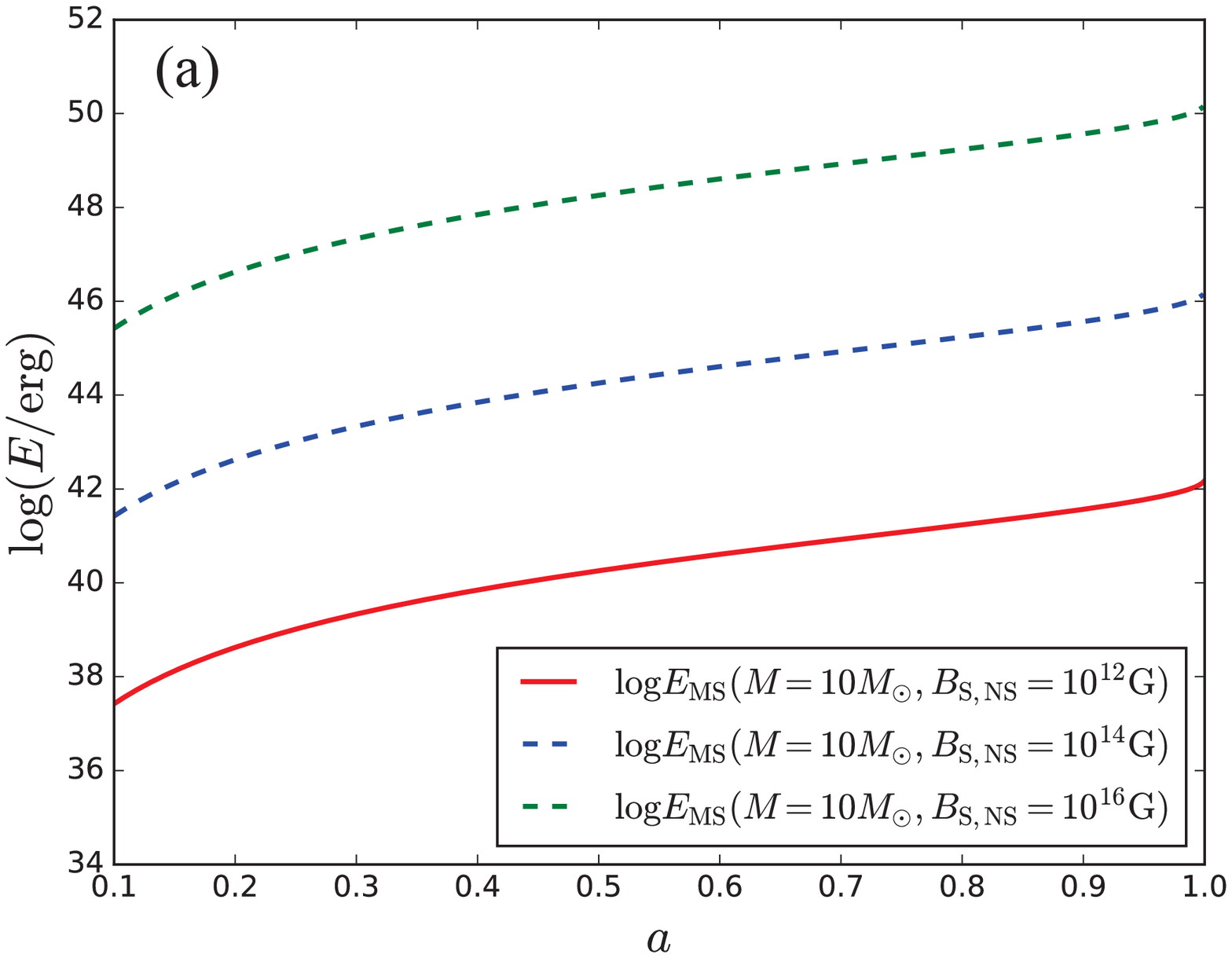}
\includegraphics[width=0.4\textwidth, height=0.18\textheight, angle=0]{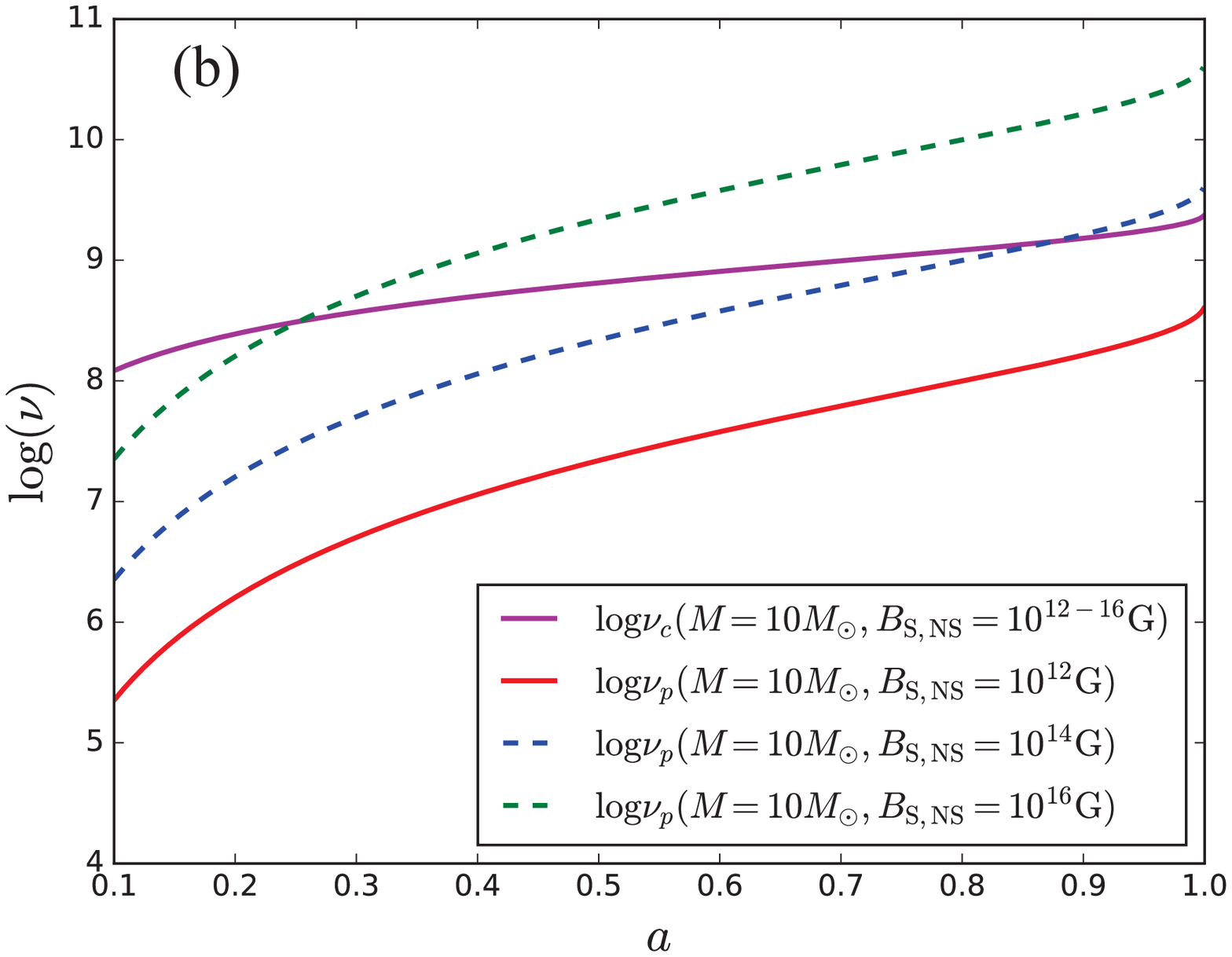}
\includegraphics[width=0.4\textwidth, height=0.18\textheight, angle=0]{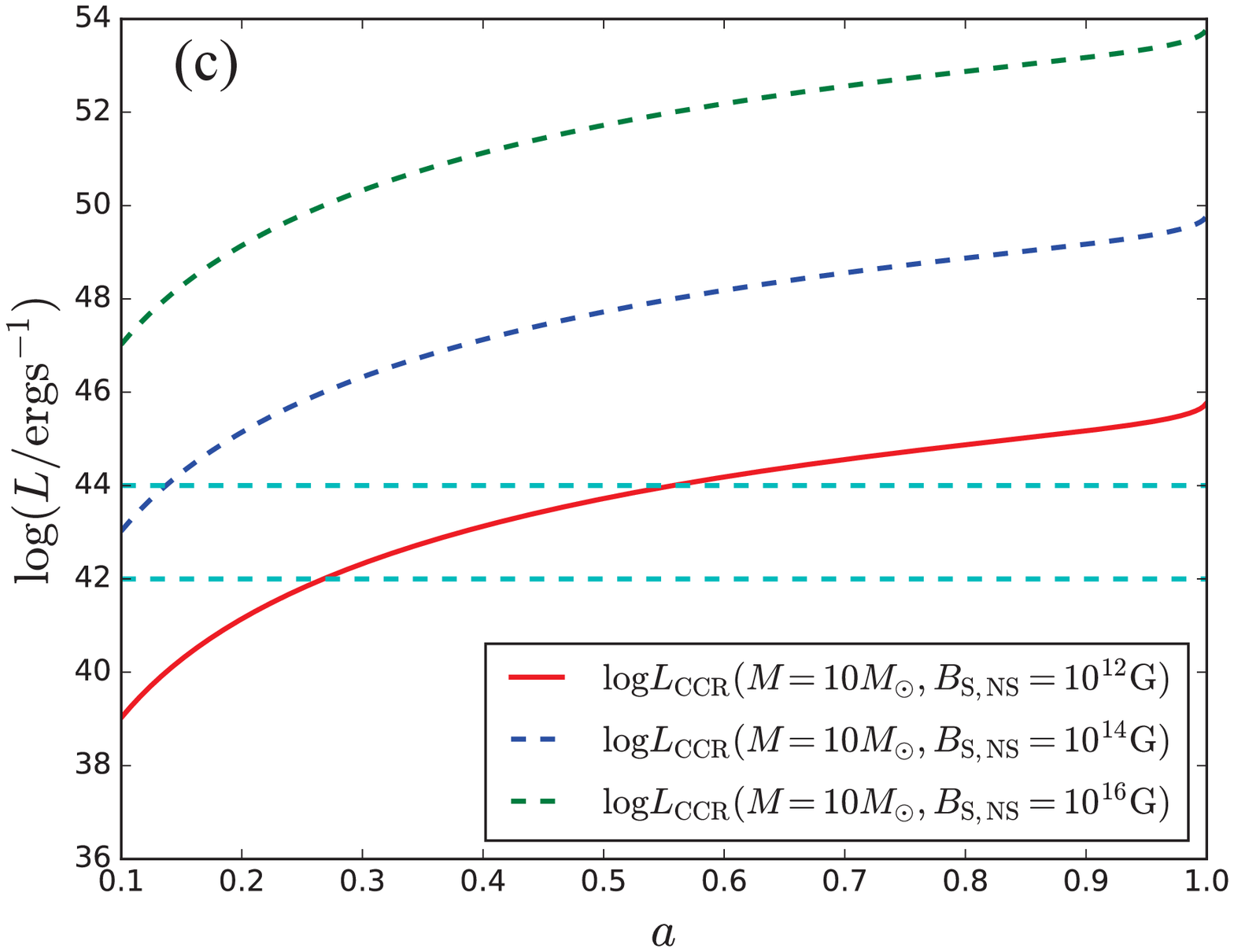}
\includegraphics[width=0.4\textwidth, height=0.18\textheight, angle=0]{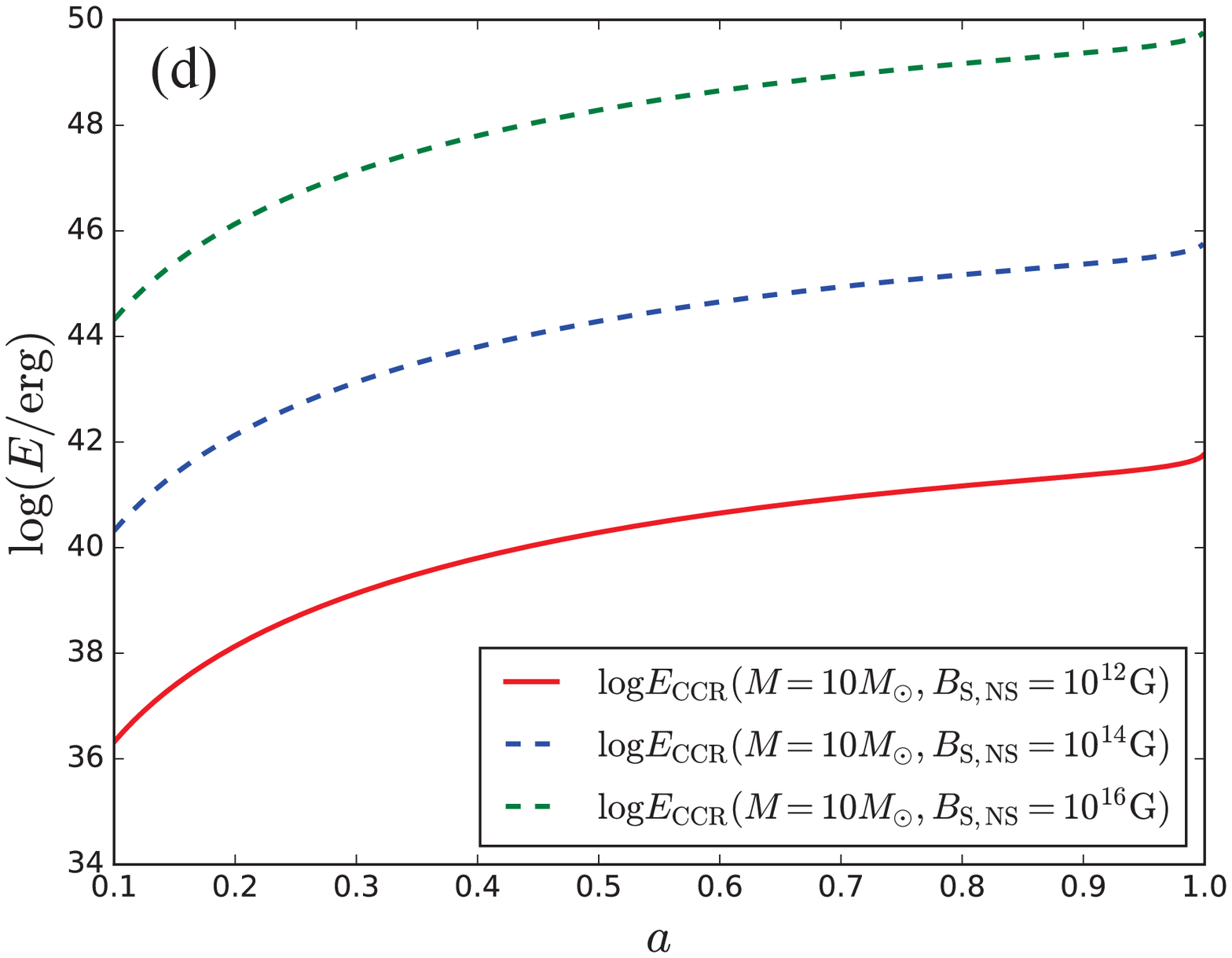}
\includegraphics[width=0.4\textwidth, height=0.18\textheight, angle=0]{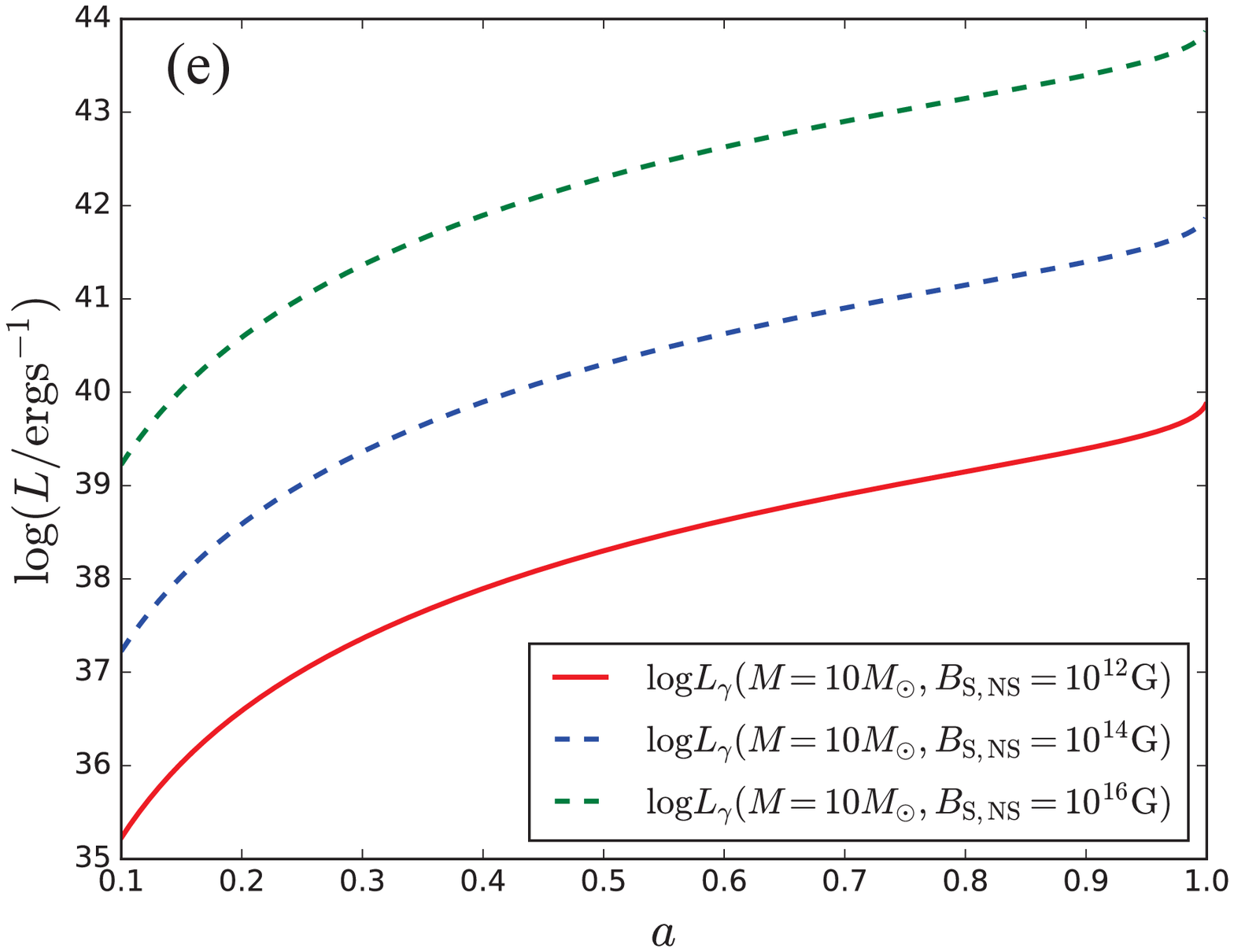}
\caption{Panel {\bf a}: The total magnetic energy in the zone of closed magnetic field lines varies with the BH spin parameter for different $B_{\rm S,NS}$. Panel {\bf b}: Comparison between the characteristic curvature radiation frequency $\nu_c$ for a relativistic charge with $\gamma\sim100$ and the plasma frequency $\nu_{\rm p}$ for different $B_{\rm S,NS}$. Panels {\bf c} and {\bf d}: The total coherent curvature radiation luminosity and energy for electrons with $\gamma\sim100$ resulted from the magnetic shock as a function of BH spin parameter for different $B_{\rm S,NS}$. The {\em cyan} lines in Panel c represent the range of a typical FRB luminosity. Panel {\bf e}: The total curvature radiation luminosity if electrons have $\gamma\sim10^7$.}
\label{fig:MS-a}
\end{figure}

After the NS is swallowed, the spinning charged BH would discharge in a force-free approximation in an time interval $\gtrsim1~{\rm ms}$ and then the entire BH magnetic field should detach and reconnect outside the horizon. This could generate a strong magnetic shock wave sweeping up the magnetosphere plasma at the speed of light.
If the magnetic shock wave energy mainly arises from the magnetic energy in the zone of closed magnetic field lines, the total magnetic energy can be estimated by
\be
E_{\rm MS}&=&\int_{R_{\rm H}}^{R_{\rm LC}}\frac{B_{\rm W}^2}{8 \pi}4\pi r^2 dr-2E_{\rm REC} \nonumber\\
&\approx& \frac{G^6a^4M^6B_{\rm S,NS}^2}{3c^{12}}\left(\frac{2}{R_{\rm H}^3}-\frac{1}{R_{\rm H}^2R_{\rm LC}}\right),
\label{eq:E_tot_MS}
\ee
which rises with increasing the NS's surface magnetic field and the BH's spin, shown in Panel a of Figure \ref{fig:MS-a}.

\cite{fr2014} suggested that curvature radiation of the shock-accelerated electrons is possibly involved in the entire emission since processes like synchrotron radiation or self-absorption are not applicable because the gyro radius of the electrons are so small that
the electrons will move essentially along the magnetic field lines. Accordingly, we consider curvature radiation in our scenario.

For the electron acceleration in the BH's magnetosphere postmerger, there are some approaches such as BH's electric field, compared to the voltage drop accelerating electrons across magnetic-field lines connecting the NS to the BH premerger in \cite{mcw2011} and \cite{do2016}.
The curvature radiation power for an electron with Lorentz factor $\gamma$ is
\be
P_{e}=2 \gamma^{4} e^{2} c / 3 R_{\mathrm{LC}}^{2},
\label{eq:P_e}
\ee
its corresponding characteristic frequency
\be
\nu_c =\frac{3 c \gamma^{3}}{4 \pi R_{\rm LC}}\approx 7 \gamma^{3}\left(\frac{R_{\rm LC}}{10^{6}\,\mathrm{cm}}\right)^{-1} \mathrm{kHz}.
\label{eq:nu}
\ee
Due to the aforementioned BH magnetosphere size $R_{\rm LC}\sim 3\times10^6-6\times10^7~{\rm cm}$,
the duration of the shock wave sweeping up the magnetosphere would be $\sim0.1-2~{\rm ms}$,
which coincides with the typical duration of an FRB. Additionally, for a relativistic electron accelerated to a Lorentz factor $\gamma\sim100$, the characteristic curvature radiation frequency is $\nu_c\sim{\rm GHz}$, falling into the FRB' frequency range.
Based on Eq. (\ref{eq:n_e}), the average electron number density can be approximated by $\bar{n}_e\approx\frac{\int^{R_{\rm LC}}_{R_{\rm H}}4\pi n_e r^2 dr}{V_{\rm LC}}$, where $V_{\rm LC}=\frac{4}{3}\pi (R_{\rm LC}^3-R_{\rm H}^3)\approx\frac{4}{3}\pi R_{\rm LC}^3$
is the volume of the whole BH magnetosphere.
As long as the NS's surface magnetic field $B_{\rm S,NS}<10^{14}~{\rm G}$, the curvature radiation frequency of an electron with $\gamma\sim100$ would be well above the surrounding plasma frequency
$\nu_{\mathrm{p}}=\gamma^{-3/2}(4\pi \bar{n}_{e}e^{2}/m_{e})^{1/2}$, as shown in the Panel b of Figure \ref{fig:MS-a}.

According to the analysis of \cite{fr2014},
the characteristic length of curvature radiation is a few $10^7$ cm, comparable with the size of shock wave---the magnetosphere size if the Lorentz factor of charges $\gamma\sim1$.
Under this condition, the entire emission would be coherent. However, for electrons with higher Lorentz factor (e.g., $\gamma\sim100$), their coherence length would become much smaller than the magnetosphere size, so the coherently emitting region should become slices and the slice number is estimated by $N_{\rm slice}\simeq VR_{\rm LC}^{-2}(c/\nu_c)^{-1}=\gamma^3$. Thus the total emitted power becomes $P_{\rm tot}=\eta_{e}N_{\rm slice}^{-1}N_{\rm e}^{2}P_{\rm e}=\eta_{e}N_{\rm slice}^{-1}(\bar{n}_eV)^{2}P_{\rm e}$ (where $\eta_{e}$ accounts for a fraction of electrons that are accelerated to Lorentz factor $\gamma$). Therefore, the luminosity by coherent curvature radiation of electrons within the zone of closed magnetic field lines can be calculated by
\be
L_{\rm CCR}=\eta_{e}\left[\bar{n}_e(V_{\rm LC}-2V_{\rm p})\right]^{2}P_{e}\gamma^{-3},
\label{eq:L_CCR}
\ee
where $V_{\rm LC}-2V_{p}$ is the volume of the zone of closed magnetic field lines and $V_{p}=\int_{R_{\rm H}}^{R_{\rm LC}}A(r)dr
\approx \frac{\pi R_{\rm H}}{R_{\rm LC}}(R_{\rm LC}^3-R_{\rm H}^3)\approx\pi R_{\rm H}R_{\rm LC}^2$.
From Eqs. (\ref{eq:Q_W,max}), (\ref{eq:B_W}), (\ref{eq:R_LC}), (\ref{eq:Omega}), (\ref{eq:P_e}), (\ref{eq:n_e}), and (\ref{eq:L_CCR}),
the luminosity can be estimated as a function of BH spin $a$ for different NS's surface magnetic field strengths,
plotted in Panel c of Figure \ref{fig:MS-a}
(with $M=10M_{\odot}$, $\gamma=100$, and $\eta_e=0.05$).
As we can see, the luminosity rapidly rises with increasing the NS's surface magnetic field and the BH spin. Moreover, it needs a low spin regime to make the luminosity fall into the characteristic FRB luminosity
$10^{42}-10^{44}~{\rm erg~s^{-1}}$ \citep{zhang2019a} if a high surface magnetic field of the NS is assumed.

This luminosity corresponds to the energy which is written by
\be
E_{\rm CCR}\approx\frac{L_{\rm CCR}R_{\rm LC}}{c},
\label{eq:E_CCR}
\ee
which also increases with increasing NS's surface magnetic field and BH's spin, illustrated in Panel d of Figure \ref{fig:MS-a}. In comparison to the result in Panel a of Figure \ref{fig:MS-a}, the energy dissipated via coherent curvature radiation of electrons is a significant fraction  of the total magnetic energy in the zone of closed magnetic field lines.

If electrons are accelerated to a very large Lorentz factor, e.g., $\gamma\sim10^7$ as argued in \cite{do2016}, their corresponding
characteristic curvature radiation frequency would fall into GeV gamma-rays based on Eq. (\ref{eq:nu}). In this case, the curvature radiation is impossibly coherent, so the total power can be estimated by
\be
L_{\rm \gamma}=\eta_e\bar{n}_e(V_{\rm LC}-2V_{\rm p})P_{e}.
\label{eq:L_gamma}
\ee
Adopting $\gamma=10^7$ and $\eta_e=1$, the luminosity as a function of BH spin for different NS's surface magnetic fields is found in Panel e of Figure \ref{fig:MS-a}. As we can see, this luminosity is very small compared to the luminosity of coherent curvature radiation for electrons with Lorentz factor $\sim100$, for the same fiducial parameter values. Moreover, it is noted that just a small fraction of ($\eta_e=0.05$) electrons with $\gamma=100$ has been assumed to account for a bright radio emission and that almost all ($\eta_e=1$) electrons with $\gamma=10^7$ only account for a faint high-energy transient. Accordingly, the magnetic shock sweeping up the magnetosphere plasma is more likely to generate a bright millisecond radio emission, e.g., an FRB, if the BH spin and the NS's surface magnetic field lie in appropriate regimes, rather than a high-energy (e.g., $\gamma$-ray) transient. Therefore, this mechanism associated with a plunging BH-NS merger event might be responsible for a fraction of non-repeating FRBs.

\subsection{BZ Mechanism}
\label{subsec:BZ Mechanism}

\begin{figure}
\plotone{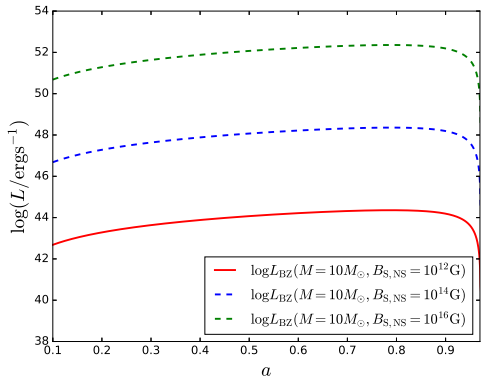}
\caption{The BZ luminosity as a function of BH dimensionless spin parameter for different $B_{\rm S,NS}$.}
\label{fig:BZ-a}
\end{figure}

During the period of magnetic reconnection at BH's poles and BH discharge resulting in a magnetic shock, the BZ mechanism might operate to extract the BH's rotational energy since the charged BH has created its own vicinal magnetosphere pervaded with charged particles postmerger.
According to \cite{do2016}, the BZ mechanism luminosity can be estimated by
\be
L_{\rm BZ}\sim \frac{\phi^2}{4\pi c}\left(\frac{ac}{R_{\rm H}}\right)^2,
\label{eq:L_BZ}
\ee
where $\phi$ is the remaining open magnetic flux of the NS after the open magnetic field lines of the NS reconnect with those of the BH, which can be approximated by
\be
\phi\sim2\pi(B_{\rm NS}-B_{\rm H})R_{\rm NS}^2{\rm sin}^{-1}\left(\frac{R_{\rm NS}\Omega_{\rm orb}}{c}\right),
\label{eq:phi}
\ee
where $B_{\rm H}=\frac{2G^3}{c^6}a^2M^3B_{\rm S,NS}R_{\rm H}^{-3}$ is the magnetic field strength at the BH's horizon from Eq. (\ref{eq:B_W}) and $\Omega_{\rm orb}$ is the orbital angular velocity near the merger time \citep{do2016}. Combining Eqs. (\ref{eq:R_H}), (\ref{eq:L_BZ}), and (\ref{eq:phi}), we obtain the BZ luminosity as a function of BH spin $a$ for the NS's surface magnetic fields $B_{\rm S,NS}=10^{12}, 10^{14}$, and $10^{16}~{\rm G}$, with BH mass $M=10M_{\odot}$, NS's radius $R_{\rm NS}=10^6~{\rm cm}$, and $\frac{2\pi}{\Omega_{\rm orb}}=1~{\rm ms}$,
as shown in Figure \ref{fig:BZ-a}. It shows that the BZ luminosity is comparable to that of Fig. 6 in \cite{do2016}.
However, there are some differences. For example, the luminosity is not sensitive to the BH spin because the magnetic field at the BH's horizon resulting from the Wald charge is very small compared to the corresponding NS's surface magnetic field until $a\sim0.7$. Only when the spin is extremely high $a>0.9$, the magnetic field strength at the BH's horizon is comparable to the NS's surface magnetic field strength, so most of the open magnetic field lines of the NS can be reconnected by those of the BH, leading to the BZ luminosity that decays as the BH spin increases.
This highly bright BZ luminosity could create a relativistic Poynting-flux-dominated jet
along the rotational axis, which has a duration similar to the timescale of magnetic reconnection and BH discharge (i.e., $\sim1~$ ms).
This jet would produce a high-energy (e.g., X-ray/$\gamma$-ray) transient due to internal magnetic dissipation in the optically thin region far away from the central engine \citep{lyu2003,zhangyan2011,ben2017,xiao2017,xiao2018,xiao2019}.

\section{Discussion and Summary}
\label{sec:discussion and Summary}
In a spinning massive BH-magnetized NS merger, the spinning BH could be charged to the maximal Wald charge quantity premerger in an electro-vacuum approximation. During the NS plunging into the BH, if the spinning charged BH could transit from electro-vacuum
to force-free cases and discharge rapidly. During the BH's discharge,
the magnetic reconnection, magnetic shock, and BZ process would probably occur,
so that they might generate a millisecond luminous EM signal, a millisecond radio bight emission like an FRB,
or a millisecond high-energy (X-ray/$\gamma$-ray) transient. Although the three processes have distinct energy sources, it is unclear  whether they would occur simultaneously. In addition, we see that if the BH has an extremely high spin $a>0.9$, the BZ luminosity decreases with $a$. These mechanisms of generating post-merger EM emission
should be also applicable to the merger of a spinning BH-charged BH binary,
in which the spinning BH immersed in the magnetic field of the charged BH could be also charged via the
Wald charging scenario at the final inspiral stage premerger.

In principle, there could be a time delay between the pre-merger emission and the post-merger emission,
at least the sum of the time from the BH's ISCO to the BH-NS merger and the light crossing time of the BH theoretically,
e.g., $\gtrsim0.5$ ms for BH mass $M=10M_{\odot}$ \citep{ming2015}.
Thus, to resolve the post-merger FRBs, it requires at least the 0.5 ms resolution of a telescope.
Fortunately, it is easy to identify an FRB possibly resulted from a pre-merger inspiral or a post-merger process if the FRB
arrives at an observer later than the GW emission during the merger.
On the other hand, it is impossible to distinguish a post-merger high energy transient
from a pre-merger emission suggested by \cite{dai2019} observationally,
if the delay between a pre-merger emission and a post-merger emission is as short as $\lesssim1$\,ms,
compared with a possible long time delay between GW-GRB association
(the time of launching a relativistic jet + the time for the jet to break out + the time for the jet to reach a GRB emission site)
\citep{zhang2019b},
e.g., the pre-merger magnetic dipole radiation process in \cite{zhang2016} can also explain the 0.4 s time delay for the candidate GW150914/GW150914-GBM association \citep{con2016,con2018,gre2016}. However, it can be seen that two post-merger mechanisms, i.e., magnetic reconnection and magnetic shock discussed in this paper are nearly independent of the NS properties such as the mass and radius, except for its surface magnetic dipole field, which is different from the pre-merger mechanisms shown in \cite{dai2019}.

If FRBs produced from the magnetic shock are not beamed
and the plunging BH-NS merger rate density upper bound $\Re_{\rm BH-NS}\lesssim610~{\rm Gpc^{-3}yr^{-1}}$ \citep{ligo2018}, then the GW-FRB association rate density from plunging BH-NS
mergers may be $\Re_{\rm GW-FRB}\lesssim610~{\rm Gpc^{-3}yr^{-1}}$.

Finally, a different mechanism to produce EM emission during BH's discharge post a spinning massive BH-magnetized NS merger
was investigated by \cite{py2019}, in which the electric energy could be released as high-energy gamma-rays
through a cascade process in analogy to the mechanism of generating $\gamma$-ray emission in a radio pulsar \citep[see Fig. 2.2 in][]{bes2004},
for which primary electrons and positrons are accelerated
in the BH electric field and then interact with
surrounding soft photons created from cyclotron radiation by non-relativistic electrons with the Goldreich-Julian density
\citep{gj1969}
via inverse Compton scattering, and then resultant hard photons colliding with soft
photons (or $\gamma+{\bf\rm \vec{E}/\vec{B}}\rightarrow e^{+}+e^{-}$) will produce more and more energetic electrons and positrons as well as high-energy photons that would escape out of the magnetosphere.

\acknowledgments
This work was supported by the National Key
Research and Development Program of China (grant No.
2017YFA0402600) and the National Natural Science Foundation
of China (grant Nos. 11573014 and 11833003).



\end{document}